\documentclass[aps, prl, 10pt, letterpaper, twocolumn, superscriptaddress]{revtex4-1}
\pdfoutput=1
\usepackage{times,xcolor,graphicx,bm}

\usepackage[pdfpagelabels,pdftex,bookmarks,breaklinks]{hyperref}
\definecolor{darkblue}{RGB}{0,0,127} 
\definecolor{darkgreen}{RGB}{0,150,0}
\hypersetup{
	colorlinks, 
	linkcolor=darkblue, 
	citecolor=darkgreen, 
	filecolor=red, 
	urlcolor=blue,
	pdftitle={Ultrahigh Error Threshold for Surface Codes with Biased Noise}, 
	pdfauthor={David Tuckett, Stephen Bartlett, Steven Flammia}
}

\begin{document}

\title{Ultrahigh Error Threshold for Surface Codes with Biased Noise}

\author{David K. Tuckett}
\author{Stephen D. Bartlett}
\affiliation{Centre for Engineered Quantum Systems, School of Physics, The University of Sydney, Sydney, NSW 2006, Australia}
\author{Steven T. Flammia}
\affiliation{Centre for Engineered Quantum Systems, School of Physics, The University of Sydney, Sydney, NSW 2006, Australia}
\affiliation{Center for Theoretical Physics, Massachusetts Institute of Technology, Cambridge, Massachusetts 02139, USA}

\date{15 December 2017}

\begin{abstract}
We show that a simple modification of the surface code can exhibit an enormous gain in the error correction threshold for a noise model in which Pauli $Z$ errors occur more frequently than $X$ or $Y$ errors. 
Such biased noise, where dephasing dominates, is ubiquitous in many quantum architectures. 
In the limit of pure dephasing noise we find a threshold of $43.7(1)\%$ using a tensor network decoder proposed by Bravyi, Suchara and Vargo. 
The threshold remains surprisingly large in the regime of realistic noise bias ratios, for example $28.2(2)\%$ at a bias of 10.
The performance is, in fact, at or near the hashing bound for all values of the bias. 
The modified surface code still uses only weight-4 stabilizers on a square lattice, but merely requires measuring products of $Y$ instead of $Z$ around the faces, as this doubles the number of useful syndrome bits associated with the dominant $Z$ errors. 
Our results demonstrate that large efficiency gains can be found by appropriately tailoring codes and decoders to realistic noise models, even under the locality constraints of topological codes. 
\end{abstract}

\maketitle

For quantum computing to be possible, fragile quantum information must be protected from errors by encoding it in a suitable quantum error correcting code. 
The surface code~\cite{Bravyi1998} (and related topological stabilizer codes~\cite{Terhal2015}) are quite remarkable among the diverse range of quantum error correcting codes in their ability to protect quantum information against local noise. 
Topological codes can have surprisingly large \emph{error thresholds}---the break-even error rate below which errors can be corrected with arbitrarily high probability---despite using stabilizers that act on only a small number of neighboring qubits~\cite{Dennis2002}. 
It is the combination of these high error thresholds and local stabilizers that make topological codes, and the surface code in particular, popular choices for many quantum computing architectures. 

Here we demonstrate a significant increase in the error threshold for a surface code when the noise is \emph{biased}, i.e., when one Pauli error occurs at a higher rate than others.
For qubits defined by nondegenerate energy levels with a Hamiltonian proportional to $Z$, the noise model is typically described by a dephasing ($Z$-error) rate that is much greater than the rates for relaxation and other energy-nonpreserving errors. 
Such biased noise is common in many quantum architectures, including superconducting qubits~\cite{Aliferis2009}, quantum dots~\cite{Shulman2012}, and trapped ions~\cite{Nigg2014}, among others. 
The increased error threshold is achieved by tailoring the standard surface code stabilizers to the noise in an extremely simple way and by employing a decoder that accounts for correlations in the error syndrome. 
In particular, using the tensor network decoder of Bravyi, Suchara and Vargo (BSV)~\cite{Bravyi2014}, we give evidence that the error correction threshold of this tailored surface code with pure $Z$ noise is $p_{c} = 43.7(1)\%$, a fourfold increase over the optimal surface code threshold for pure $Z$ noise of $10.9\%$~\cite{Bravyi2014}. 

These gains result from the following simple observations.
For a $Z$ error in the standard formulation of the surface code, the stabilizers consisting of products of $Z$ around each plaquette of the square lattice contribute no useful syndrome information. 
Exchanging these $Z$-type stabilizers with products of $Y$ around each plaquette still results in a valid quantum surface code, since these $Y$-type stabilizers will commute with the original $X$-type stabilizers. 
But now there are twice as many bits of syndrome information about the $Z$ errors.
Taking advantage of these extra syndrome bits requires an optimized decoder that can use the correlations between the two syndrome types. 
The standard decoder based on minimum-weight matching breaks down at this point, but  the BSV decoder is specifically designed to handle such correlations. 
We show that the parameter $\chi$, which defines the scale of correlation in the BSV decoder, needs to be large to achieve optimal decoding, so in that sense accounting for these correlations is actually necessary. 
These two ideas---doubling the number of useful syndrome bits and a decoder that makes optimal use of them---give an intuition that captures the essential reason for the increased threshold.  It is nonetheless remarkable just how large an effect this simple change makes. 

\begin{figure}[tbp!]
\begin{center}
\includegraphics{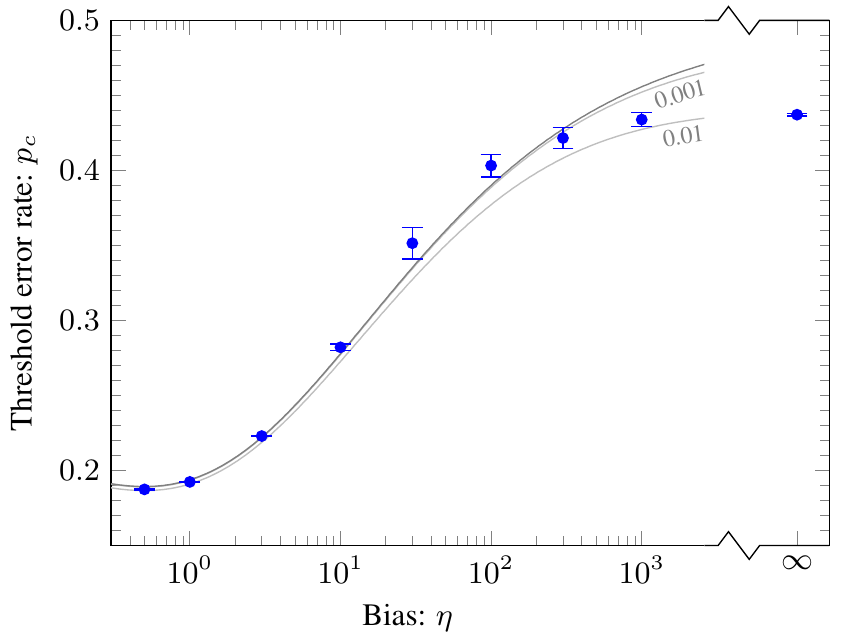}
\caption{Threshold error rate $p_c$ as a function of bias $\eta$. 
The dark gray line is the zero-rate hashing bound for the associated Pauli error channel. 
Lighter gray lines show the hashing bound for rates $R = 0.001$ and $0.01$ for comparison; the surface code family has rate $1/n$ for $n$ qubits. 
Blue points show the estimates for the threshold using the fitting procedure described in the main text together with 1-standard-deviation error bars.
The point at the largest bias value corresponds to infinite bias, i.e., only $Z$ errors.}
\label{fig:threshold-v-bias}
\end{center}
\end{figure}

We also consider more general Pauli error models, where $Z$ errors occur more frequently than $X$ and $Y$ errors with a nonzero bias ratio of the error rates. 
We show that the tailored surface code exhibits these significant gains in the error threshold even for modest error biases in physically relevant regimes: for biases of $10$ (meaning dephasing errors occur 10 times more frequently than all other errors), the error threshold is already $28.2(2)\%$.
Figure~\ref{fig:threshold-v-bias} presents our main result of the threshold scaling as a function of bias.
Notably, we find that the tailored surface code together with the BSV decoder performs near the hashing bound for all values of the bias.

\begin{figure}[tb!]
\begin{center}
\includegraphics{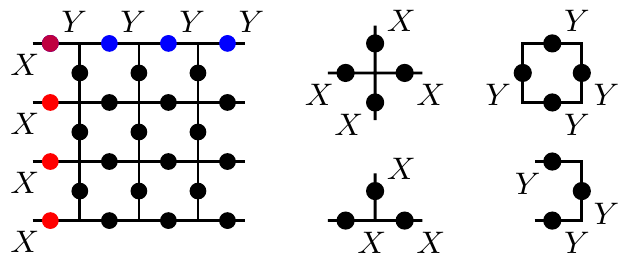}
\caption{The modified surface code, tailored for biased $Z$ noise, with logical operators given by a product of $Y$ along the top edge and a product of $X$ along the left edge. 
The stabilizers are shown at right.}
\label{fig:BSV}
\end{center}
\end{figure}

\paragraph{Error correction with the surface code.---}
The surface code~\cite{Bravyi1998} is defined by a 2D square lattice having qubits on the edges with a set of local stabilizer generators. 
In the usual prescription, for each vertex (or plaquette), the stabilizer consists of the product of the $X$ (or $Z$) operators acting on the neighboring edges. 
We simply exchange the roles of $Z$ and $Y$, as shown in Fig.~\ref{fig:BSV}. 
By choosing appropriate ``rough'' and ``smooth'' boundary conditions along the vertical and horizontal edges, the code space encodes one logical qubit into the joint $+1$ eigenspace of all the commuting stabilizers with a code distance $d$ given by the linear size of the lattice. 

A large effort has been devoted to understanding error correction of the surface code and the closely related toric code~\cite{Kitaev2003}. 
The majority of this effort has focused on the cases of either pure $Z$ noise, or depolarizing noise where $X$, $Y$, and $Z$ errors happen with equal probability; see Refs.~\cite{Terhal2015, Brown2015} for recent literature reviews. 
Once a noise model is fixed, one must define a decoder, and the most popular choice is based on minimum-weight matching (MWM). 
This decoder treats $X$ and $Z$ noise independently, and it has an error threshold of around $10.3\%$ for pure $Z$ noise with a naive implementation~\cite{Dennis2002, Wang2003}, or $10.6\%$ with some further optimization~\cite{Stace2010}. 
Many other decoders have been proposed, however, and these are judged according to their various strengths and weaknesses, including the threshold error rate, the logical failure rate below threshold, robustness to measurement errors (fault tolerance), speed, and parallelizability. 
Of particular note are the decoders of Refs.~\cite{Duclos-Cianci2010, Duclos-Cianci2014, Fowler2013, Wootton2012, Delfosse2014, Hutter2014, Torlai2017, Baireuther2017, Krastanov2017}, since these either can handle, or can be modified to handle, correlations beyond the paradigm of independent $X$ and $Z$ errors.

\paragraph{The BSV decoder.---}
Our choice of the BSV decoder~\cite{Bravyi2014} is motivated by the fact that it gives an efficient approximation to the optimal maximum likelihood (ML) decoder, which maximizes the \emph{a posteriori} probability of a given logical error conditioned on an observed syndrome. 
This decoder has also previously been used to do nearly optimal decoding of depolarizing noise~\cite{Bravyi2014}, achieving an error threshold close to estimates from statistical physics arguments that the threshold should be $18.9\%$~\cite{Bombin2012}. 
[In fact, our own estimate of the depolarizing threshold using the BSV decoder is $18.7(1)\%$.] 
Because it approximates the ML decoder, the BSV decoder is a natural choice for finding the maximum value of the threshold for biased noise models. 

The decoder works by defining a tensor network with local tensors associated with the qubits and stabilizers of the code. 
The geometry of the tensor network respects the geometry of the code. 
Each index on the local tensors has dimension 2 initially, but during the contraction sequence, this dimension grows until it is bounded by $\chi$, called the bond dimension. 
When $\chi$ is exponentially large in $n$, the number of physical qubits, then the contraction value of the tensor network returns the exact probabilities conditioned on the syndrome of each of the four logical error classes. 
Such an implementation would be highly inefficient, but using a truncation procedure during the tensor contraction allows one to work with any fixed value of $\chi \ge 2$ with a polynomial runtime of $O(n \chi^3)$. 
In this way, the algorithm provides an efficient and tunable approximation of the exact ML decoder, and in practice small values of $\chi$ were observed to work well~\cite{Bravyi2014}. 
We refer the reader to Ref.~\cite{Bravyi2014} for the full details of this decoder.

\paragraph{Biased Pauli error model.---} 
A Pauli error channel is defined by an array $\bm{p}=(1-p,p_x,p_y,p_z)$ corresponding to the probabilities for each Pauli operator $I$ (no error), $X$, $Y$, and $Z$, respectively. 
We define $p = p_x+p_y+p_z$ to be the probability of any single-qubit error, and we always consider the case of independent, identically distributed noise. 
We define the bias $\eta$ to be the ratio of the probability of a $Z$ error occurring to the total probability of a non-$Z$ Pauli error occurring, so that $\eta = p_z/(p_x+p_y)$.
For simplicity, we consider the special case $p_x = p_y$ in what follows. 
Then for total error probability $p$, $Z$ errors occur with probability $p_z = [\eta/(\eta + 1)]$, and $p_x = p_y = [1/2(\eta + 1)]p$.
When $\eta=1/2$, this gives the standard depolarizing channel with probability $p/3$ for each nontrivial Pauli error, and taking the limit $\eta \to \infty$ gives only $Z$ errors with probability $p$.
Biased Pauli error models have been considered by a number of authors~\cite{Aliferis2008, Aliferis2009, Rothlisberger2012, Napp2013, Brooks2013a, Webster2015, Robertson2017}, but we note that there are several different conventions for the definition of bias. 
Comparison between channels with different bias but the same total error rate is facilitated by the fact that the channel fidelity to the identity is a function only of $p$.

\paragraph{Hashing bound.---}
The quantum capacity is the maximum achievable rate at which one can transmit quantum information through a noisy channel~\cite{Wilde2013}. 
The hashing bound~\cite{Lloyd1997, Shor2002, Devetak2005} is an achievable rate which is generally less than the quantum capacity~\cite{DiVincenzo1998}. 
For Pauli error channels, the hashing bound takes a particularly simple form~\cite{Wilde2013} and says that there exist quantum stabilizer codes that achieve a rate $R = 1 - H(\bm{p})$, with $H$ being the Shannon entropy. 
The proof of achievability involves using random codes, and it is generally hard to find explicit codes and decoders that perform at or above this rate for an arbitrary channel, especially if one wishes to impose additional constraints such as local stabilizers. 
The quantum capacity itself is still unknown for any Pauli channel where at least two of $(p_x,p_y,p_z)$ are nonzero.


\begin{figure}[tbp!]
  \begin{center}
    \includegraphics{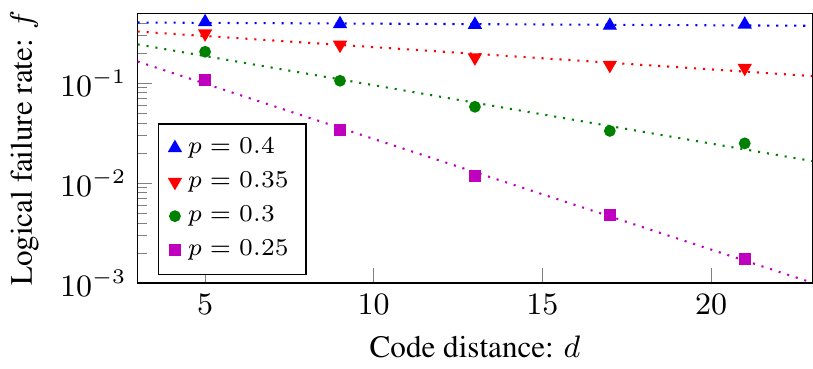}
    \caption{Exponential decay of the logical failure rate $f$ with respect to code distance $d$ in the regime $p < p_c$ for $\eta = 100$ and $\chi = 48$. 
      We observe scaling behavior of the form $f \sim \exp(-\alpha d)$ where $\alpha$ depends on the bias and is an increasing function of $(p_c-p)$.
      In this bias regime, the decoder performance is likely farthest from optimal, but the decay is still clearly exponential over this range. 
      Other values of $\eta$ show the same general scaling behavior, though with different decay rates $\alpha$. 
      The statistical error bars from 30\,000 trials per point are smaller than the individual plot points in every case.}
    \label{fig:expdecay}
  \end{center}
\end{figure}

\paragraph{Numerics.---}
Our numerical implementation makes only a minor modification to the BSV decoder.  To avoid changing the definitions of the tensors used in Ref.~\cite{Bravyi2014}, we use the symmetry by which we can exchange the role of $Z$ noise in the modified surface code with the role of $Y$ noise in the standard surface code.
Then all of the definitions in Ref.~\cite{Bravyi2014} carry over unchanged. 
The only difference is that we perform two tensor network contractions for each decoding sequence. 
There is an arbitrary choice as to whether to contract the network row-wise or column-wise. 
Rather than pick just one, we average the values of both contractions. 
We empirically observe improved performance with this modification. 

\begin{figure}
  \begin{center}
    \includegraphics{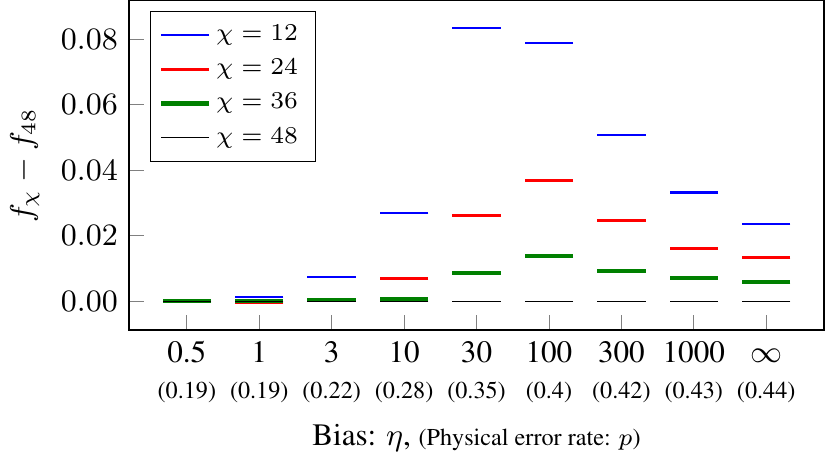}
    \caption{Convergence of the decoder as a function of $\chi$ near the threshold for distance $d=21$.
      We observe that the logical failure rates $f_\chi$ stabilize with increasing $\chi$ for both low and high biases.
      However, in the intermediate bias regime $f_\chi$ is still decreasing noticeably between increments of $\chi$, suggesting that $\chi > 48$ would be required for a good approximation to the optimal ML decoder.}
    \label{fig:convergence-v-bias}
  \end{center}
\end{figure}

For each value of the bias $\eta \in \{$0.5, 1, 3, 10, 30, 100, 300, 1000, $\infty\}$, we estimate the logical failure rate $f$ using the BSV decoder to obtain the sample mean failure rate on 30\,000 random trials for a selection of physical error rates $p$ in the region near the threshold $p_c$ for code distances $d \in \{$9, 13, 17, 21$\}$.
We use a rather large value of the bond dimension $\chi$ for our simulations, specifically $\chi = 48$, although for bias $\eta < 30$ we already observe that the decoder converges well with $\chi = 36$.
However, we still do not observe complete convergence of the decoder at $\chi = 48$ in the regime of intermediate bias around $\eta = 100$.
The decoder convergence with $\chi$ is displayed in Fig.~\ref{fig:convergence-v-bias}, which shows the estimate of the logical failure rate for the $d=21$ code near the threshold.
Performance of the decoder and convergence with $\chi$ generally improve as bias increases again beyond $\eta = 300$, but it is likely that further improvements are possible in the intermediate bias regime. Although the decoder at $\chi = 48$ is not achieving an optimal failure rate in the intermediate regime, we see excellent convergence for most of the range of bias and across the full range of bias we observe threshold behavior.
Moreover, this threshold is at the hashing bound for all $\eta \le 100$.
In the regions that are a fixed distance below the threshold, as in Fig.~\ref{fig:expdecay}, we observe an exponential decay in the logical failure rate $f \sim \exp(-\alpha d)$, where $\alpha$ may depend on the bias and is an increasing function of $(p_c - p)$. 
This constitutes strong evidence of an error correction threshold.

We note that $\chi = 48$ was the largest used in our simulations, so we do not know if the saturation of the decoder performance for bias $\eta \ge 300$ is a real effect, or a side effect of having too small a value of $\chi$.
Although we observe convergence and threshold behavior, we do not know how much the performance might improve for larger values of $\chi$ since, as seen in Fig.~\ref{fig:convergence-v-bias}, there is apparently still some room for improvement.
It is possible that the saturation is a real effect, however, since even at infinite bias there are still logical errors of weight $2d = O(\sqrt{n})$ that consist only of $Z$ errors.
This is in contrast to the classical repetition code, which has a threshold of $50\%$ and a distance $O(n)$.
One possibility to address this is to use a surface code with side lengths $L \times W$, where $L$ and $W$ are relatively prime, for example just choosing $W=L+1$.
We empirically observe that the $Z$-distance (i.e., the distance when restricted only to $Z$ errors) of the code scales like $O(n)$ for this modification of the surface code.
In fact, on a toric code with $L$ and $W$ both odd and relatively prime, the $Z$-distance is provably $O(n)$~\cite{Bravyi2017}.
These observations are currently being explored, and will be addressed in more detail in forthcoming work.

\begin{figure*}
  \begin{center}
    \includegraphics{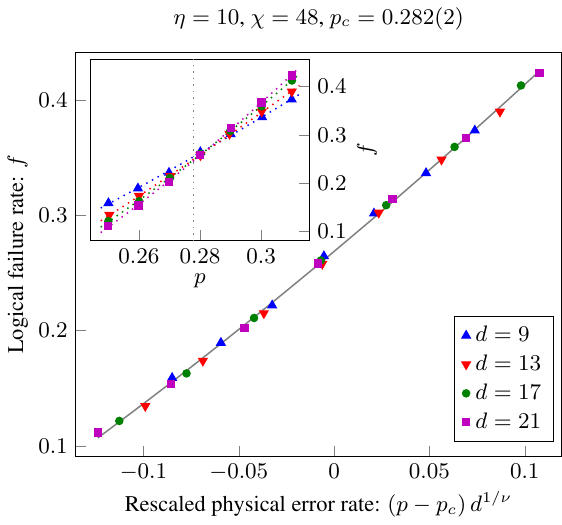}
    \enskip\includegraphics{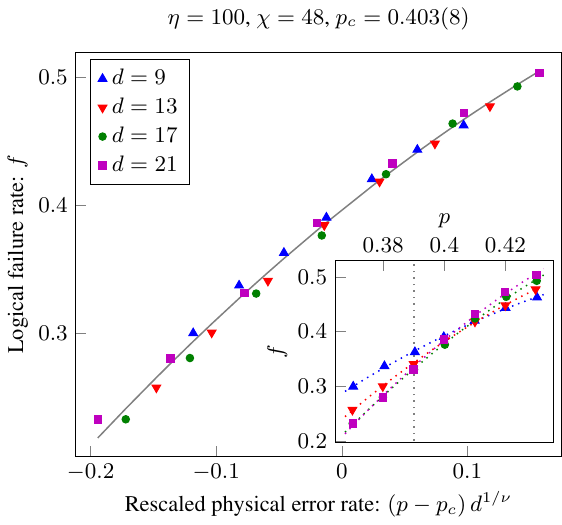}
    \enskip\includegraphics{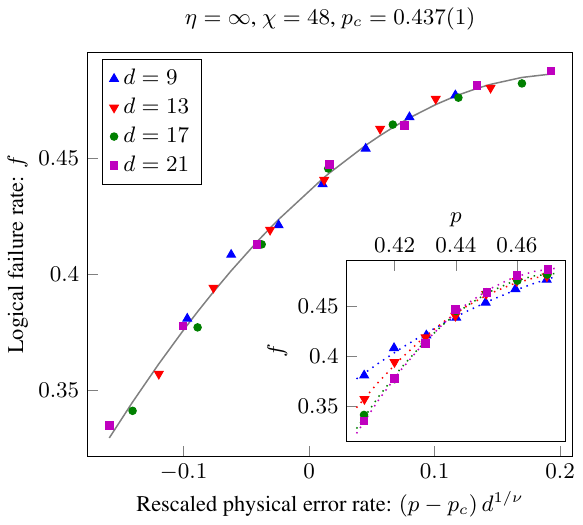}
    \caption{Logical failure rate $f$ as a function of the rescaled error rate $x = (p-p_c) d^{1/\nu}$ for biases $\eta\in\{$10, 100, $\infty\}$. 
      The solid line is the best fit to the model $f = A + Bx+Cx^2$. 
      The insets show the raw sample means over 30\,000 runs for various values of $p$, and the dotted gray vertical line indicates the hashing bound. 
      Even for the case of $\eta = 100$ where the decoder performance was likely furthest from optimal we still see good agreement with the fit model.}
    \label{fig:fit}
  \end{center}
\end{figure*}

To obtain an explicit estimate of the threshold $p_c$, we use the critical exponent method of Ref.~\cite{Wang2003}. 
If we define a correlation length $\xi = (p-p_c)^{-\nu}$ for some critical exponent $\nu$, then in the regime where $d \gg \xi$ we expect that the behavior of the code is scale invariant. 
In this regime, since the code distance $d$ corresponds to a physical length, the failure probability should depend only on the dimensionless ratio $d/\xi$, a conjecture that was first empirically verified in Ref.~\cite{Wang2003}.  
This suggests defining a rescaled variable $x = (d/\xi)^{1/\nu} = (p-p_c) d^{1/\nu}$ so that the failure rate expanded as a power series in $x$ is explicitly scale invariant at the critical point $p_c$ corresponding to $x=0$. 
It is then natural to consider a model for the failure rate given by a truncated Taylor expansion in the neighborhood around $p_c$. 
We use a quadratic model, $f = A + B x + C x^2$, and then fit to this model to find $p_c, \nu$ and the nuisance parameters $A, B, C$. 
A discussion on the limits of the validity of this universal scaling hypothesis can be found in Ref.~\cite{Watson2014a}. 
We plot our estimates of $f$ for various values of $p$ and $d$ for the representative cases of $\eta \in \{$10, 100, $\infty\}$ in Fig.~\ref{fig:fit} together with rescaled data as a function of $x$.
A visual inspection confirms good qualitative agreement with the model. 

The critical exponents method gives precise estimates of $p_c$ with low statistical uncertainty.
However, systematic biases might affect the accuracy of the estimate and must be accounted for.
Finite-size effects typically cause threshold estimates to decrease as larger and larger code distances are added to the estimate.
Additionally, the suboptimality of the decoder due to small $\chi$ values in the intermediate bias regime may have overestimated each individual logical failure rate.
This latter effect does not directly imply that we have also overestimated the threshold $p_c$, and the data remain consistent with the fit model in spite of this as can be seen in Fig.~\ref{fig:fit}.
On balance, we expect that our estimates might decrease somewhat in the intermediate bias regime.
Our final error bars were obtained by jackknife resampling, i.e.\ by computing, for each fixed $\eta$, the spread in estimates for $p_c$ when rerunning the fit procedure with a single distance $d$ removed, for each choice of $d$.
Our results are summarized in Fig.~\ref{fig:threshold-v-bias}.

\paragraph{Fault tolerant syndrome extraction.---}
Our study has focused on the error correction threshold under the assumption of ideal syndrome extraction.
To see if the gains observed in this setting carry over to applications in fault-tolerant quantum computing, one would need to consider the effects of faulty syndrome measurements and gates.
A full fault-tolerant analysis is beyond the scope of this work, but we briefly consider the key issues here.

First, the BSV decoder that we have used to investigate this ultrahigh error threshold is not fault tolerant, but some clustering decoders are~\cite{Duclos-Cianci2014}.
Developing efficient, practical fault-tolerant decoders with the highest achievable thresholds remains a significant challenge for the field.

An added complication with a biased noise model is that the gates that perform the syndrome extraction must at least approximately preserve the noise bias in order to maintain an advantage~\cite{Aliferis2009}.
For the tailored surface code studied here, one could appeal to the techniques of Refs.~\cite{Aliferis2009,Brooks2013a}, where we note that $Y$-type syndromes can be measured using a minor modification of the $X$-syndrome measurement scheme.
We note that these syndrome extraction circuits are significantly more complex (involving the use of both ancilla cat states and gate teleportation) compared with the standard approach for the surface code with unbiased noise, and this added complexity will undoubtedly reduce the threshold.

More optimistically, we note that the standard method for syndrome extraction in the surface code~\cite{Fowler2012} can be directly adapted to this tailored code and maintains biased noise on the data qubits.
Ancilla qubits are placed in the centers of both the plaquette and vertex stabilizers of Fig.~\ref{fig:BSV}, and they will be both initialized and measured in the $X$ basis.
Sequences of controlled-$X$ (vertex) and controlled-$Y$ (plaquette) gates, with the ancilla as the control and data qubits as the target, yield the required syndrome measurements analogous to the standard method.
In this scheme, we note that high-rate $Z$ errors on the ancilla are never mapped to the data qubits; low-rate $X$ and $Y$ errors on the ancilla can cause errors on the data qubits but the noise remains biased.  Measurement errors will occur at the high rate, but this can be accommodated by repeated measurement.
Note that, as argued by Aliferis and Preskill~\cite{Aliferis2009}, native controlled-$X$ and controlled-$Y$ gates are perhaps not well motivated in a system with a noise bias, but nonetheless this simple scheme illustrates that, in principle, syndromes can be extracted in this code while preserving the noise bias.
To develop a full fault-tolerant syndrome extraction circuit in a noise-biased system would require a complete specification of the native gates in the system and an understanding of their associated noise models.

\paragraph{Discussion.---}
Our numerical results strongly suggest that in systems that exhibit an error bias, there are significant gains to be had for quantum error correction with codes and decoders that are tailored to exploit this bias.
It is remarkable that the tailored surface code performs at the hashing bound across a large range of biases.
This means that it is not just a good code for a particular error model, but broadly good for any local Pauli error channel once it is tailored to the specific noise bias.
It is also remarkable that a topological code, limited to local stabilizers, does so well in this regard.

Many realizations of qubits based on nondegenerate energy levels of some quantum system have a bias---often quite significant---towards dephasing ($Z$ errors) relative to energy-nonconserving errors ($X$ and $Y$ errors).
This suggests tailoring other codes, and in particular other topological codes, to have error syndromes generated by $X$- and $Y$-type stabilizers.
Even larger gains might be had by considering biased noise in qudit surface codes~\cite{Anwar2014, Watson2015}.

For qubit topological stabilizer codes, the threshold for exact ML decoding with general Pauli noise can be determined using the techniques of Ref.~\cite{Bombin2012}, which mapped the ML decoder's threshold to a phase transition in a pair of coupled random-bond Ising models.
It would be interesting to explore this phase boundary for general Pauli noise beyond the depolarizing channel that was studied numerically in Ref.~\cite{Bombin2012}.

We have employed the BSV decoder to obtain our threshold estimates because of its near-optimal performance, but it is not the most efficient or practical decoder for many purposes.
One outstanding challenge is to find good practical decoders that can work as well or nearly as well across a range of biases.
The clustering-type decoders~\cite{Duclos-Cianci2010, Duclos-Cianci2014} appear well suited for this task, and they have the added advantage that some versions of these decoders (e.g., Ref.~\cite{Bravyi2011}) generalize naturally to all Abelian anyon models such as the qudit surface codes.

The most pressing open question related to this work is whether the substantial gains observed here can be preserved in the context of fault-tolerant quantum computing.

\paragraph{Acknowledgements.---}
This work is supported by the Australian Research Council (ARC) via Centre of Excellence in Engineered Quantum Systems (EQuS) Project No. CE110001013 and Future Fellowship No. FT130101744, by the U.S. Army Research Office Grants No. W911NF-14-1-0098 and No. W911NF-14-1-0103, and by the Sydney Informatics Hub for access to high-performance computing resources.

\end{document}